\documentclass[preprint,12pt]{elsarticle}

\usepackage{epsfig}

\usepackage{amssymb}

\usepackage{lineno}

\journal{Nuclear Instruments and Methods}

\begin{document}

\begin{frontmatter}



\title{Thermomechanical design of a static gas target for electron accelerators}


\author[label1]{B. Brajuskovic}
\author[label1] {T. O'Connor}
\author[label1] {R. J. Holt}
\author[label1,label3] {J.~Reneker}
\author[label2] {D.~Meekins}
\author[label2] {P.~Solvignon}

\address[label1]{Argonne National Laboratory, Argonne, IL 60439}
\address[label2]{Jefferson Laboratory, Newport News, VA 23606}
\fntext[label3]{Present address: Sandia National Laboratories, Albuquerque, NM 87185}
\begin{abstract}
Gas targets are often used at accelerator facilities. A design of  high-pressure gas cells that are suitable for hydrogen and helium isotopes at relatively high electron beam currents is presented.  In particular, we consider rare gas targets, $^3$H$_2$ and $^3$He.  In the design, heat transfer and mechanical integrity of the target cell are emphasized.  ANSYS 12 was used for the thermo-mechanical studies of the target cell.  Since the ultimate goal in this study was to design a gas target for use at the Jefferson Laboratory (JLab), particular attention is given to the typical operating conditions found there.  It is demonstrated that an aluminum alloy cell can meet the required design goals.

\end{abstract}


\end{frontmatter}


\section{Introduction}
\label{intro}
Since the electromagnetic interaction is relatively weak, a relatively large luminosity is necessary for electron scattering experiments.  The problem addressed here is developing an experiment design that makes optimal use of rare target gases such as $^3$H$_2$ and $^3$He.  The presence of tritium gas places numerous stringent requirements on the target cell.  Recently, two proposals \cite{Petratos:2010,Solvignon:2011} that make use of a tritium target have been assigned high priority and recommended for approval by the Jefferson Lab Program Advisory Committee.  Tritium targets have been used at a number of electron accelerators: the Stanford High Energy Physics Laboratory \cite{Hughes:1966zza}, the MIT-Bates Electron Accelerator Center \cite{Beck:1985yp,Beck:1989bi}, and the Saclay Electron Accelerator Lab \cite{Juster:1985sd}.  The general characteristics of these targets are given in Table 1.

\begin{table}[ht]
\caption{Characteristics of tritium targets that have been used at electron accelerators.  The last entry represents properties of a possible target for Jefferson Lab.}
\label{tab:a}
\begin{tabular}{lcrcrr}
\hline
\hline

  Lab & Year & Quantity & Thickness & Current & Luminosity  \\
   &  & (PBq)/(kCi) &  (g/cm$^2$)   &     ($\mu$A)   &   ($\mu$A-g/cm$^2$) \\

\hline

Stanford  & 1966  &  0.93/25  &  0.8  &  1  & 0.8 \\
MIT-Bates & 1985 & 6.7/180 &  0.3   & 20  & 6.0 \\
Saclay  & 1985  & 0.37/10  &  1.2  &  10  & 12.0 \\
JLab    &  -   &  0.04/ 1.17 &  0.1 & 20  & 2.0  \\

\hline
\hline
\end{tabular}
\end{table}

The targets that we consider in this study are hydrogen, deuterium and tritium gas targets that are 10 bar in pressure at room temperature and 40 cm long.  A $^3$He target of the same length but at 20 bar is also considered. Both stainless steel and aluminum target cells were studied.  An electron beam in the GeV region and a current of 20 $\mu$A with a rastered spot size of 3 mm in diameter was assumed for this study. The beam was assumed to be uniform over the entire spot size.  With these assumptions, the characteristics of this target are compared with previous targets in Table 1 in the entry under JLab.  In terms of luminosity this target is competitive with previous targets, but makes use of the smallest inventory of tritium, only 43.3 TBq (1.17 kCi), of the previous targets.

There have been studies of static and circulating gas targets at ion beam facilities.  In particular a study of the reduction in the gas target density was performed for a few- mm diameter proton or heavy-ion beam incident on static and circulating hydrogen gas targets.  It was found \cite{Goerres, Yamaguchi} that the threshold in beam power per unit length in the gas to induce density fluctuations is 10 mW/mm. For the hydrogenic targets considered here and a 20~$\mu$A electron beam, the linear power density is approximately 9.5 mW/mm, while that for the $^3$He target is approximately 19 mW/mm.  According to Gorres {\it et al.} \cite{Goerres} this would result in a negligible reduction in the hydrogenic target thicknesses and about a 10\% reduction in the $^3$He target thickness. Of course, the density correction factors can be determined from beam current scans during the actual experiment.  Unlike low-energy proton or heavy-ion beams, the high-energy electron beam loses a relatively small amount of energy in the thin windows and target gas.  It is essentially a nearly constant energy loss over the length of the target cell.  At Jefferson Lab in particular, the beam shape, diameter and current are quite stable.  The actual beam has a diameter between 50 and 100 $\mu$m.  The beam is rastered to a size of about 3 mm on the target and thus has essentially a rectangular spatial distribution.  The effects of density fluctuations due to beam-induced target gas heating can be measured periodically during the experiment, say with each electron beam energy change, by measuring the target scattering rate as a function of the electron beam current.

\section{Mechanical design and cooling}

The objective is to design high-pressure static gas cells that contain hydrogenic and helium gases that can be used as a target at a relatively high beam current at an electron accelerator facility.  Since one of the gases is tritium, a large safety factor is necessary.
The design parameters are demanding and sometimes conflicting.

The design criteria for the target were:
\begin{itemize}
\item The amount of rare gas should be optimized.
\item The target cell material must be compatible with tritium storage.
\item Any possibilities for leaks (seals) should be minimized.
\item The target cell must reside in vacuum.
\item A reasonably high density of target gas should be achieved with a cell less than 400 mm in length and a diameter of at least 12.5 mm.
\item Electron beam entry, exit and side windows should be no thicker than 0.46 mm of Al to preserve a reasonable signal to background ratio based on a GEANT4 simulation.
\item At least four target cells plus an empty cell for background measurements should be provided.
\item With a 20 $\mu$A electron beam, the warmest part of the cells with hydrogenic gases should not exceed 180~K, the threshold for electron beam-induced corrosion of aluminum.
\item The cell should have a high safety factor and survive off-normal operation such as loss of coolant or beam raster for much longer than the time for an interlock to turn off the beam.
\end{itemize}

After realizing that the thermo-mechanical properties of stainless steel were not sufficient for this application, we investigated
 aluminum 2219-T851 because Al is compatible with tritium usage \cite{handbook} and because this high strength alloy is weldable.  The use of welded joints rather than flanges with seals, would reduce the potential for possible leaks as well as permit accurate heat transfer simulations.
When developing the mechanical design for this target, the emphasis was placed on minimizing the
material in the path of the beam (the endcaps of the cell) while maintaining a high factor of safety.  Further consideration was given to minimizing the thickness of the side walls of the cell through which scattered particles must traverse.

\begin{figure}[ht]
\begin{center}
\includegraphics[angle=-90,width=.9\textwidth]{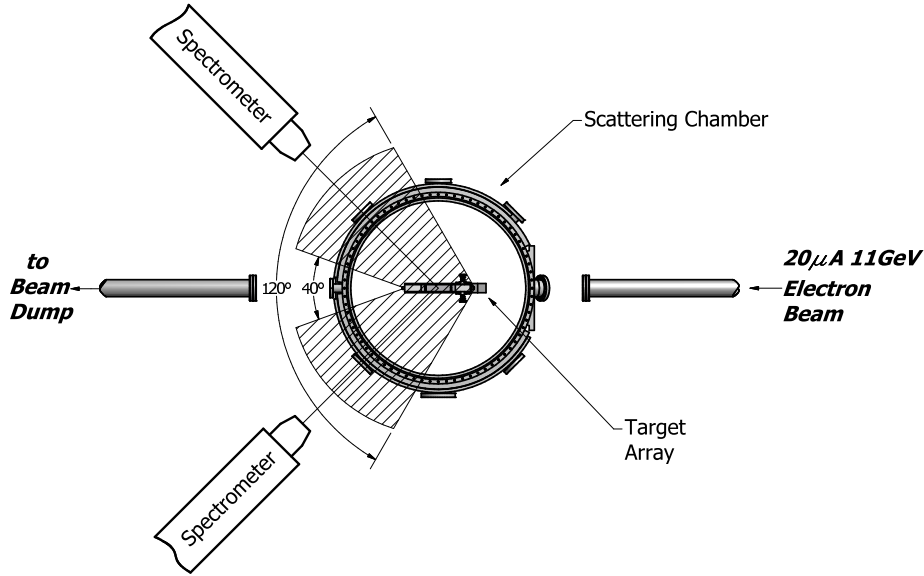}
\caption{Schematic diagram of the top view of the experiment indicating the location of the target cells with respect to the beam and spectrometer.  The evacuated scattering chamber also serves as secondary containment for the tritium target.
\label{fig:top}}
\end{center}
\end{figure}

\begin{figure}[ht]
\begin{center}
\includegraphics[angle=-90,width=0.65\textwidth]{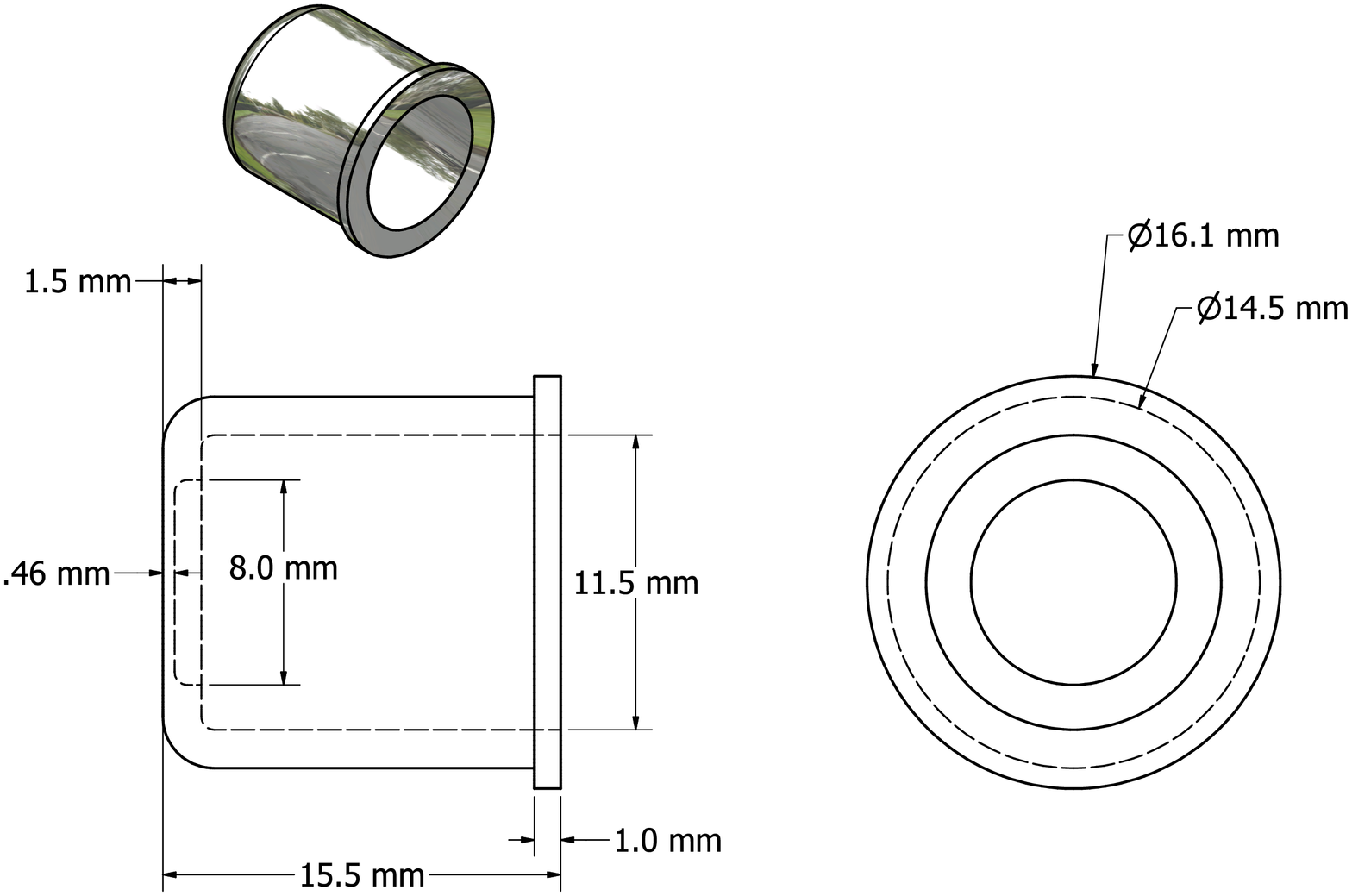}
\caption{Detail of the endcap design for the high pressure gas targets.
 Note the thin ``windows'' for the incident and exiting electron beam.
\label{fig:endca}}
\end{center}
\end{figure}

A top view of overall setup of the target, electron beam and spectrometers is given in Fig.~\ref{fig:top}.
The target cell assembly is located inside an evacuated scattering chamber as indicated in the figure.  The scattering chamber is completely isolated from the beam line and serves as secondary containment for the tritium gas target cell.
The most vulnerable components in the target cell assembly are the endcaps where the primary electron beam enters and exits the target. It is desirable to
keep these as thin as possible but when the beam is running they will experience the most heating in the
central portion. Since the rastered beam has a diameter of 3~mm, we designed an endcap with a thin
central diameter of 8~mm. Outside this area the thickness was increased significantly in order to optimize heat conduction. Given the mechanical and heat transfer properties of Aluminum 2219-T851, a central thickness
of 0.46~mm, increasing to 1.5~mm outside of this area, was found to be acceptable. A flange was included in
the design in order to facilitate welding and the overall height of the endcap was chosen to be 11.5~mm in
order to place the welds far away from the heat generated in the central portion as indicated in Fig.~\ref{fig:endca}.  Although Al 2219-T851 can be age hardened after welding, the material strength does not return fully to the pre-welded conditions.

\begin{figure}[ht]
\begin{center}
\includegraphics[angle=-90,width=0.80\textwidth]{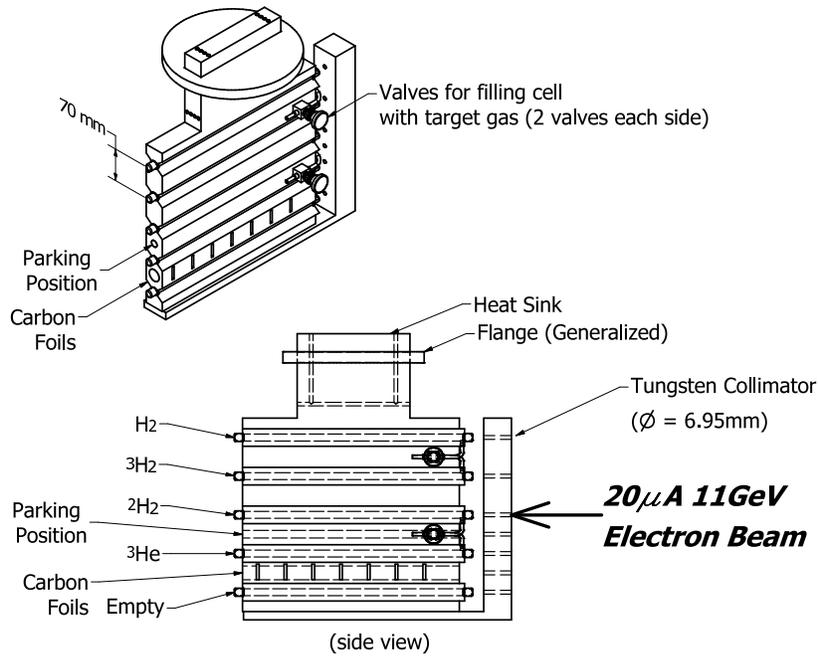}
\caption{Diagram of all five target gas cells coupled to a heat sink.
\label{fig:cell}}
\end{center}
\end{figure}

For this application five target cells are necessary, one for each of the gases of hydrogen, deuterium, tritium and helium-3 and one empty cell for background measurements.
The main bodies of all five target cells will be machined out of a single piece of aluminum. The
maximum thickness will be approximately 41~mm. Along the axis of each target cell a wedge of material
will be removed on each side in order to reduce the wall thickness within the field of view as shown in Fig.~\ref{fig:cell}.
Use of a single piece will allow the individual targets to be passively cooled. A large heat sink block,
integral to the target array, will be actively cooled. The ends of the targets will contain counterbores
designed to fit the flanges on the endcaps and provide access for e-beam welding.  A hole will be drilled into the thick portion of the target array. This will provide a ``parking
position'', as indicated in the figure, and allow running the beam for background studies without any targets or heat generation.  This position is distinct from the empty target cell since the empty target cell will be used to measure background scattering from the cell entrance and exit windows.
Thermal analysis shows that this configuration will prevent the individual target cells and endcaps from
overheating. The hydrogenic targets which have the most stringent temperature requirements will be placed nearest to the heat sink block.

The target lengths can be determined from 3-D measurements of the entire assembly on a coordinate measurement machine at various cell pressures.  Comparison between inner, outer, and cell wall dimensions can give dimensions accurate to 0.1 mm.  The target gases are expected to be greater than 99.5~\% isotopically pure.  The fill temperature and pressure in the cell will be monitored to ~0.3~\% at 1~Mpa.  The total uncertainty in the cell length is expected to be less than 2~\%.

Fig.~\ref{fig:cell} also shows the overall assembly. Four valves will be mounted to the sides of the target
array, out of the field of view, in order to fill and seal the individual cells. These are tritium-compatible
stainless-steel valves. Since the target array will be aluminum an aluminum-to-stainless
transition piece will be used to join these parts. The target cells are located 70~mm apart.  This provides sufficient space for the large azimuthal acceptance of the spectrometers over the full range of the target length.  An array of 25-mm thick tungsten collimators is attached to the
target array just upstream of the entrance window. The purpose of these collimators is to block the beam
in the event of severe beam pointing errors which have the potential to damage or even burn through our target
walls. Finally, the heat sink block will be welded to a UHV flange and the entire target array will be
placed into a vacuum chamber, with the cooling lines attached on the outside.  The targets can be changed by a lifter assembly that moves the targets as a group vertically while the electron beam remains in a fixed position.

\section{Thermal and structural analysis}
Two overall goals of the target design were to keep a large factor of safety, at least 5, for the gas cells and to keep the hottest part of the Al target cell below 180 K for the hydrogenic targets. The main reason for keeping the temperature low is that a previous study \cite{flower} found a threshold of 180 K for electron-irradiation induced aqueous corrosion of Al.
Thermal and structural analysis was performed to determine safety factors for normal operating conditions.  Analysis was performed using the ANSYS 12 Finite Element Final (FEA) Package.  In all analyzed cases the material of the target cells was Al 2219 T851. The material properties of aluminum were obtained from ref. \cite{mpdb}.  The values for thermal conductivity of hydrogen, deuterium and helium were obtained from ref. \cite{touloukian}.   The value for tritium thermal conductivity was assumed to be 60\% of that for hydrogen. However, a variation of the thermal conductivity in the range from 50\% to 200\% of that of hydrogen was performed in the simulations and a negligible effect was found.   The beam diameter was taken to be 3 mm. The wall thicknesses were 0.45~mm and 0.456~mm for end-cap windows and thin side-walls, respectively. Because the cells are to be operated in a vacuum chamber, the initial net gas pressure in the cells at room temperature was 11~bar for hydrogen, deuterium and tritium and 21~bar for helium.

The thermal analysis was performed with the following assumptions:
\begin{itemize}
\item	20 $\mu$A 11-GeV electron beam with 3 mm diameter
\item	The beam-generated heat \cite{volmer} was 6.7~W in the endcaps, 3.8~W in hydrogen gases and 7.5~W in helium gas. The heat generated by the tritium decay ($\sim$ 50~mW) was neglected in the analysis.
\item	The analysis was performed for a 77~K LN$_2$ coolant at a mass flow of 30 g/s and a coefficient of convective heat transfer of 835 W/m$^2$ as well as for cryogenic helium gas which maintained the wall of the cooling channel at 35~K.
\end{itemize}
The computed results for LN2 cooling are shown in Table 2. The results are shown for the helium target exposed to the beam that has the largest heat generation and is furthest from the cooling source, the tritium target that raises the highest safety concerns, and for the hydrogen target that is closest to the cooling channels.  In all cases the highest overall temperatures were computed in the center of the downstream end-caps of the exposed cells while the highest computed temperatures in the exposed gas were a fraction of a degree lower.
The highest temperature of 185.6 K was computed when the helium-3 cell was exposed to the beam. The highest temperature in the aluminum walls of cells of hydrogen  gases, 175.1 K (not shown in Table 2), was computed for the deuterium cell. The highest computed temperature in the tritium cell was 169.9 K while maximum computed temperature of the aluminum walls of the hydrogen cell exposed to the beam was 168.1 K.  The highest computed side wall temperature for the tritium cell exposed to the beam was 102~K. The overall highest side wall temperature of 122.2~K was computed for the helium-3 cell exposed to the beam. The maximum computed temperatures in the thin side-walls of the exposed cells were in all cases at least 60~K lower than the maximum temperatures in the end-caps. In all analyzed cases the temperature distribution in non-exposed  cells was relatively uniform and the maximum temperature of non-exposed cells remained below 112~K.

\begin{table}[!]
\caption{Calculated temperatures, maximum equivalent stresses and safety factors for LN2 cooling.  The yield stresses are given for the maximum temperatures.}
\label{tab:f}
\begin{tabular}{|l|c|c|c|c|}
\hline
\hline
  & \multicolumn{1}{c|} {Max. Temp.}
  & \multicolumn{1}{|c|} {Max. Stress}
& \multicolumn{1}{c|} {Yield Stress}
& \multicolumn{1}{c|}  {Factor of}    \\
  & \multicolumn{1}{c|}  {(K)}  &  {(MPa)}  &  {(MPa)}  &  {Safety}  \\
\hline
\multicolumn{5}{|l|} {$^3$He cell exposed to the beam}  \\
\hline

$^3$He Cell Endcaps & 185.6 & 61.1 & 373.5 & 6.1        \\
$^3$He Cell Side Walls & 122.2 & 35.3 & 394.2  & 11.2   \\
$^3$H$_2$ Cell Endcaps & 102.1 & 17.4 & 406.2 & 25.5      \\
$^3$H$_2$ Cell Side Walls & 102.6 & 26.8 & 402.4 & 15.0  \\

$^2$H$_2$ Cell Endcaps & 111.1 & 16.1 & 398.7 & 24.7       \\
$^2$H$_2$ Cell Side Walls & 111.5 & 34.4 & 398.5 & 11.6  \\

H$_2$ Cell Endcaps & 94.1 & 16.0 & 406.5 & 25.5       \\
H$_2$ Cell Side Walls & 94.3 & 52.9 & 406.2 & 7.7    \\
\hline
\multicolumn{5}{|l|} {$^3$H$_2$ cell exposed to the beam}  \\
\hline
$^3$He Cell Endcaps & 95.6 & 32.1 & 405.5 & 12.6        \\
$^3$He Cell Side Walls & 95.6 & 13.3 & 405.5 & 30.4   \\

$^3$H$_2$ Cell Endcaps & 169.9 & 44.4 & 377.7 & 8.5      \\
$^3$H$_2$ Cell Side Walls & 102.0  & 35.0 & 405.3 & 11.5  \\

$^2$H$_2$ Cell Endcaps & 96.3 & 16.8 & 405.2 & 25.0       \\
$^2$H$_2$ Cell Side Walls & 96.3 & 22.3 & 405.5 & 18.1  \\

H$_2$ Cell End-caps & 92.2 & 15.6 & 407.1 & 26.1       \\
H$_2$ Cell Side walls & 92.3 & 32.2 & 407.1 & 12.6    \\
\hline

\multicolumn{5}{|l|} {H$_2$ cell exposed to the beam}  \\
\hline
$^3$He Cell Endcaps & 89.4 & 30.0 & 408.5 & 13.6        \\
$^3$He Cell Side Walls & 89.4 & 11.1 & 408.5 & 36.7  \\

$^3$H$_2$ Cell Endcaps & 90.7 & 15.7 & 407.9 & 26.0     \\
$^3$H$_2$ Cell Side Walls & 90.7 & 20.3 & 407.9 & 20.1  \\

$^2$H$_2$ Cell Endcaps & 89.5 & 15.7 & 408.4 & 26.0       \\
$^2$H$_2$ Cell Side Walls & 89.5 & 9.7 & 408.4 & 42.3  \\

H$_2$ Cell Endcaps & 168.1 & 43.8 & 378.3 & 8.6       \\
H$_2$ Cell Side Walls & 99.6 & 21.5 & 403.7 & 18.8    \\

\hline
\end{tabular}
\end{table}

The structural analysis was performed to calculate stresses due to the thermal expansion of the structure and due to the static pressure of stored gases. The change in static pressure of the contained gases, first due to cooling to the cryo temperature and then due to heating by exposure of the target to the beam, was taken in account. The results indicate that the equivalent stresses in the walls of end-cap windows and in the thin sidewalls of the gas cells are much higher than in the rest of the target structure. In all analyzed cases the highest values of equivalent stress are computed at the center of inner side of end-cap window walls of the exposed cell. The highest value of equivalent stress of 61.2 MPa is computed for helium cell exposed to the beam. The maximum equivalent stresses in the cells of hydrogen gases exposed to the beam are in the 43.8-45.4 MPa range with the higher values computed for the cells that are further from the heat sink. It is interesting to note that for the cells that are not exposed to the beam, the maximum values of equivalent stress in end-caps are found at the center of outer side of window walls.
The maximum values of computed equivalent stresses in the side walls are lower than those computed at the center of windows in the end-caps and there is less regularity in their location. In three out of four analyzed cases the highest value of equivalent stress is computed in the side wall of the hydrogen (upper-most) cell. The highest equivalent stress value computed in the side-walls, 52.9 MPa, is found in the side-walls of hydrogen cell when the helium (lower-most) cell is exposed to the beam. The only case when the maximum equivalent stress is not computed in the hydrogen cell side-wall is when the tritium cell (second from the top) is exposed to the beam. In that case the maximum value of equivalent stress, 35 MPa, is computed in the side-wall of the exposed cell but the maximum computed equivalent stress in the side-walls of hydrogen cell is very close at 32.2 MPa.

 The lowest maximum equivalent stress in the side-wall of the hydrogen cell (21.5 MPa) is computed for the case when the hydrogen cell is exposed to the beam.  The pattern of maximum equivalent stresses strongly indicates that the stress values in the end-caps are defined by the interaction of the beam and end-cap window while the stresses in the side walls are more dependent on the deformation of the entire structure and are influenced with the overall geometry and the way that the target structure is restrained within the rest of experimental set-up.

 Values for the factor of safety which represent the yield stress-to-maximum computed stress ratio were calculated for the temperature dependent yield stress values\cite{mpdb} and in determination of the yield stress values the maximum computed  temperatures were used. The calculated values of the factor of safety are given in the last column of Table 2. The factor of safety was found to be greater than 6 for all analyzed cases. The lowest calculated value for the tritium cell was 8.5 and was calculated for the endcaps when the tritium cell was exposed to the beam.

The simulation was repeated for cryogenic helium cooling rather than LN2 cooling.  This is an interesting case since, contrary to intuition, the factors of safety are in general lower than those in the LN2-cooled targets.  The main reason for this is that thermal gradients are in general larger between the heated regions and the cooled regions because of the larger temperature difference and the corresponding thermal contractions and expansions.  The results of this simulation are summarized in Table 3.  In this case, we only list the highest cell endcap temperatures and the minimum factors of safety for several beam currents and targets.  When the beam impinges on the tritium target, it is found that the tritium target can be operated with 25 $\mu$A without exceeding the critical temperature for corrosion, but exceeds 180~K with a 30~$\mu$A beam. Although the factor of safety is reasonably good for the 30~$\mu$A case, it would still be problematic because of possible corrosion of the cell walls.  The $^3$He target can be operated up to 50 ~$\mu$A with a safety factor of approximately three.  Although the maximum window temperature of $^3$He cell reaches 283~K the temperature of the aluminum in the hydrogenic gas cells  remains below 130~K.  It is interesting to note that when the luminosity can be achieved with a thicker tritium target, say 1.6 kCi rather than the 1.17 kCi, the cell temperature remains lower and the factor of safety actually improves somewhat over gaining the luminosity by increasing the beam current.

\begin{table}[!]
\caption{Calculated maximum endcap temperatures and minimum cell safety factors when the electron beam impinges on a 1.17 kCi tritium target (first three rows), a 1.6 kCi tritium target (fourth row), and the $^3$He target (last row) for cryogenic helium cooling.}
\label{tab:g}
\begin{tabular}{|c|c|c|c|}
\hline
\hline
 \multicolumn{1}{|c|} {Target in Beam}
& \multicolumn{1}{c|} {Beam Current}
  & \multicolumn{1}{c|} {Max. Temp.}
& \multicolumn{1}{c|}  {Factor of}    \\
  & \multicolumn{1}{c|}  {($\mu$A)}  &  {(K)}  &  {Safety}  \\
\hline

 1.17 kCi $^3$H$_2$ target & 20 & 144.0 & 7.0        \\

 & 25 & 162.9 & 6.7     \\

& 30 & 181.3 & 6.6       \\
\hline

 1.6 kCi $^3$H$_2$ target& 25 & 163.8 & 7.0     \\

\hline
$^3$He target & 50 & 283.0 & 3.1      \\
\hline
\end{tabular}
\end{table}

Preliminary thermal transit simulations were performed both for a loss-of-coolant incident and for a loss of raster event.  The results indicate that the cell should survive these incidents at least until an interlock shuts the beam off, typically 50~$\mu$s. In fact, the simulations indicate that it would take 10 minutes for the cell window to reach 180~K after loss of coolant.

After having established an optimum design from the heat transfer calculation and basic considerations for electron scattering, a Monte Carlo simulation using GEANT4 was developed. As stated, an upstream W collimator was designed to protect the target from unexpected offset in the beam position and to avoid beam scraping on the wall of the target. It is expected that the beam halo scraping on the collimator would trigger the radiation monitors and shut off the beam.

\section{Summary and conclusions}
A viable design is put forward for a high-pressure static gas target suitable for use at electron accelerator facilities.  Because of the combination of high thermal conductivity, relatively high yield stress and compatibility with tritium gas, an aluminum alloy was chosen for this study.  In particular, we chose Al 2219 because of its weldability.  Our goal was to produce a design that was optimized for heat transfer so that 20~$\mu$A of electron beam could traverse the target.  Also, the target cell geometry was optimized to conserve the amount of gas necessary for a reasonable luminosity so that relatively rare gases such as $^3$H$_2$ and $^3$He could be used. Finally, in order to minimize hydrogen diffusion and possible corrosion or embrittlement of the cell, it was important to keep the operating temperature of the hydrogenic target cells below a temperature of 180~K in the presence of the electron beam.  A thermo-mechanical FEA analysis of the design confirms that these goals are achievable. It was found that with cryo-cooled helium gas rather than LN2 would permit a beam current approaching 30~$\mu$A. Presently, a target based on this general design is under development for use at Jefferson Lab.

\section*{Acknowledgments}
We wish to thank E. J. Beise, W. Korsch, G. G. Petratos, R. Ransome, R. Ricker, B. Somerday and B. Wojtsekhowski for extremely useful discussions.  We also thank P. Den Hartog and E. Trakhtenberg for substantive comments.  This work was supported by the Department of Energy, Offices of Nuclear Physics and Basic Energy Sciences under contract nos.  DE-AC02-06CH11357 and DE-AC05-060R23177.

\bibliographystyle{model1a-num-names}
\bibliography{target_fin_rev}

\end{document}